# GraphBreak: Tool for Network Community based Regulatory Medicine, Gene co-expression, Linkage Disequilibrium analysis, functional annotation and more

Abhishek Narain Singh
Web: ABioTek www.tinyurl.com/abinarain
abhishek.narain@iitdalumni.com

**Abstract**

Graph network science is becoming increasingly popular, notably in big-data perspective where understanding individual entities for individual functional roles is complex and time consuming. It is likely when a set of genes are regulated by a set of genetic variants, the genes set is recruited for a common or related functional purpose. Grouping and extracting communities from network of associations becomes critical to understand system complexity, thus prioritizing genes for disease and functional associations. Workload is reduced when studying entities one at a time. For this, we present GraphBreak, a suite of tools for community detection application, such as for gene co-expression, protein interaction, regulation network, etc.Although developed for use case of eQTLs regulatory genomic network community study- results shown with our analysis with sample eQTL data- Graphbreak can be deployed for other studies if input data has been fed in requisite format, including but not limited to gene co-expression networks, protein-protein interaction network, signaling pathway and metabolic network. GraphBreak showed critical use case value in its downstream analysis for disease association of communities detected. If all independent steps of community detection and analysis are a step-by-step sub-part of the algorithm, GraphBreak can be considered a new algorithm for community based functional characterization. Combination of various algorithmic implementation modules into a single script for this purpose illustrates GraphBreak's novelty. Compared to other similar tools, with GraphBreak we can better detect communities with overrepresentation of its member genes for statistical association with diseases, therefore target genes which can be prioritized for drug-positioning or drug-repositioning as the case be.

**Introduction**

By signaling pathway networks, bio-based chemical molecular networks- such as protein–protein interaction networks, gene co-expression networks, gene regulatory networks, metabolic networks- provide a graphical representation of cellular and tissue systems. To guide biological experiments, networks must be analyzed by means of community detections, if every gene transcript, gene product, protein, genotypic variants, such as SNPs, were to be characterized individually they could be very difficult to perform. To increase understanding of network phenomena, network scientists deploy models. For this, we introduce GraphBreak, a network analysis tool to understand variant based regulation of expression of genes as a community. It conducts community analysis of overrepresentation association with disease and other function phenotypes, along with linkage association of SNP variants for causative prioritization. This network-based community detection has been examined in detail from genomic variants, such as SNPs with expression of genes, but GraphBreak can be used for any network analysis, such as those discussed above.For example, WGCNA [Langfelder, P., et. al, 2008] has been widely used for gene co-expression analysis. Compared to WGCNA, where a weight is assigned for any connections between two genes,GraphBreak assumes a weight of 1 for every connection. Condor, a similar tool for community detection in gene regulation context,[Platig, 2016] is used for the purpose that models' network assume it to be a bipartite graph.

**Motivation for developing a new community detection tool**

GraphBreak was inspired by Condor [Platig, 2016], which implements modularity detection in bipartite network [Barber, Michael, 2008]. Regulatory connections between genomic variants and associated gene expressed is bipartite by default, given that there were no connections between any two genomic variants (such as SNPs), nor any connections between any two genes expressed. As simple community detection in a connected graph is performant in obtaining the communities, there was no need to classify community detection problem as a bipartite graph if data itself is arranged in such a way. For this, our approach was to find an implementation module for various community detection algorithms with preferably parallel computing capability, which can then be used for our own purpose, such as for regulatory genomic network analysis.

**Existing algorithms for community detection influencing GraphBreak development**

Sets of nodes are partitioned by a class of network analysis methods into subsets depending on graph structure. For instance, all nodes in a connected component are reachable from each other. Using breadth-first search, a network's connected components can be computed in linear time. Community detection is the task of identifying groups of nodes in the network that are significantly more densely connected among another than to the rest of nodes. Various definitions of the structure to be discovered- community- is a data mining problem. NP-hard optimization problem defines this task by community quality measures-first and foremost modularity [Girvan and Newman, 2002]. The goal of GraphBreak was for end users to choose from popular community detection algorithms, such as Louvain's algorithm [Blondel et al., 2008] and Girvan-Newman algorithm [Girvan and Newman, 2002].

NetworKit [Staudt and Meyerhenke, 2015] was chosen for community detection because with parallel script in place, we should be ready in the future for high volume of data. NetworKit is a Python API (application programming interface) that approaches community detection from the perspective of modularity maximization and engineer parallel heuristics, which deliver a good balance between solution quality and running time. NetworKit API has been previously used for solving biological network problems such as protein-protein interaction (PPI) network [Flick, 2014], therefore reassuring this API's deployment for biochemical application.

PLP algorithm implements community detection by extracting communities from a labeling of the node set, or label propagation [Raghavan et al., 2007]. Community detection with Louvain method [Blondel et al., 2008] can be classified as a locally greedy, bottom-up multilevel algorithm. NetworKit authors recommend, PLM algorithm with optional refinement step as the default choice for modularity-driven community detection in large networks. PLP delivers a better time to solution for very large network in the range of billions of edges, but with a qualitatively different solution and worse modularity. Therefore Girvan-Newman and Louvain algorithm for community detection using Networkx [Hagberg et al., 2008] module in Python and the parallel version of Louvain algorithm (PLM) using the NetworKit module in Python was implemented, as we felt it was illogical to use PLP to save time but compromise modularity solutions obtained, as for the time being our data sizes are not in the order of terabytes that would require us to see the problem from a really 'Big Data' perspective.

NetworkX has functionalities to help convert the graph data from one format to another. NetworkX [Hagberg et al., 2008], a de-facto standard for the analysis of small to medium networks in a Python environment, is not suitable for massive networks due to its pure Python implementations. Even though our data size was not of the category very 'Big Data', it was large enough to be classified as 'Big Data', so using algorithmic implementations in NetworkX for community detection was clearly time and resource demanding. Girvan-Newman and Louvain algorithm were implemented with the help of NetworkX Python module for small and medium size data,we strongly emphasize use of GraphBreak by parallel implementation of Louvain's algorithm (PLM) for large-scale data, which we developed using NetworKit Python module. Data sizes are currently not extremely large, for practicality purposes, so the choice of deploying Louvain algorithm or Girvan-Newman will not have a detrimental effect on execution time.

**GraphBreak PseudoCode and Workflow**

Flowchart for GraphBreak workflow is shown in Figure C, followed by pseudocode.

**Materials used**

GTEx [GTEx consortium, 2013] V7 data was used for analysis, cis-eQTLs of which was generated by FastQTL[Ongen et al., 2016] by GTEx consortium as the following downloadable file [GTEx_Analysis_v7_eQTL.tar.gz](GTEx_Analysis_v7_eQTL.tar.gz) and [GTEx_Analysis_v8_eQTL.tar.gz](GTEx_Analysis_v8_eQTL.tar.gz) from the following GTEx Portal URL: https://www.gtexportal.org/home/datasets. We took only sample data from artery – aorta from GTEx V7 and artery-coronary tissue from GTEx V8 as a test case given the significance of this tissue in coronary artery disease. GraphBreak and Condor were executed to detect communities and see the efficacy by downstream analysis of genes in each of communities for their statistical significance with disease. The computing facility from 'Bioinformatics Centre at University of Eastern Finland', Kuopio campus was used.

**Method: Execution of GraphBreak and Condor for community detection**

GraphBreak was used to plot connections between variants and genes. Figure 1. below shows a sample GraphBreak plot for 3 genes as an example.

Connections between genomic variants and genes were plotted using Condor as in Figure 2.

Community detection methods were called once connections were plotted into memory of data structures. Correspondences with Condor authors lead to them developing community plots and 16 communities were detected by Condor as shown in figure 3.

GraphBreak obtained about 56 communities comprising of expressed genes and associated genomic variants for each, GraphBreak plots for 5 communities with only genes, shown in figure 4.

Plotting 57 communities, using GraphBreak would have made the plot non-aesthetic, so a few were plotted to ensure that we received expected result. Sum of the number of genes in each of the communities obtained from Condor and GraphBreak were plotted. Condor does not use Louvain algorithm,but igraph, for the purpose of finding the communities [Barber, Michael, 2008], another method which implements modularity detection in bipartite network. Figure 5 and Figure 6 show the sum of the genes in each community obtained from Condor and GraphBreak, respectively.

Each community obtained by Condor and GraphBreak had genomic variants, such as SNPs with them, which can be plotted in a similar way. For our current illustration purpose, we focus only on set of genes derived for each of the communities as downstream analysis.

**Downstream Analysis Methods**

**Method 1: Over-Representation analysis of Genes obtained for each of the communities**

Set of genes obtained from communities by Condor or GraphBreak can be tested for their significance of association for functionalities and diseases by means of over-representation analysis as detailed in the following paper [Sorin Drăghici et. al.], high performance implementation of which was published here [Antonio Fabregat et. al.] are based on collection of knowledge bases from experimental findings. Reacfoam[Antonio Fabregat et al., 2018]improvised the graphical representation that uses 'Voronoid Diagram' to depict statistically significant associations for over-representation analysis with functionalities and diseases and uses open source knowledge of biochemical pathways. Figure 7 shows statistically significant results for one community's (1061) set of genes being significantly associated with diseases, metabolism of proteins, immune system, metabolism, and cell cycle.

**Method 2: LD Prioritization analysis of genomic variants obtained for each of the communities**

Using existing database for variant classifications, such as RegulomeDB [Boyle et al., 2012], which classifies genomic variants based on already known properties, known as an eQTL, or having transcription factor binding site, DNase peak, etc.,genomic variants grouped in each community can then be annotated for functionalities. Next, variants can be evaluated for LD amongst each other to reduce possible causative variants from those short-listed in step one. Figure 8 below shows a simple LD association plot obtained by LDMatrix web-application-part of LDlink package [Machiela et. al., 2015].

**Benchmarking**

Igraph[Csardi and Nepusz, 2006] and graph-tool's [Peixoto, 2015]functionality and target use cases are similar to NetworKit, confirmed by authors of NetworKit. Packaged as Python modules, they provide a broad feature set for network analysis workflows and active user communities. They address scalability issues by implementing core data structures and algorithms in C or C++; graph-tool builds on Boost Graph Library and parallelizes some kernels using OpenMP.

Gephi [Bastian et al., 2009], a GUI application for Java platform with a strong focus on visual network exploration, is geared towards network science but differs in important aspects from NetworKit. Pajek [Batagelj and Mrvar, 2004], a proprietary GUI application for Windows operating systemoffers visualization features-with analysis capabilities like NetworKit. PajekXXL, a variant, uses less memory and focuses on large datasets. SNAP [Leskovec and Sosič, 2014] network analy-sis package has recently adopted the hybrid approach of C++ core and Python interface.

Related efforts from algorithm engineering community are KDT [Lugowski et al., 2012] (algebraic built, distributed parallel backend), GraphCT [Ediger et al., 2013] (focused on massive multithreading architectures likeCray XMT), STINGER (a dynamic graph data structure with some analysis capabilities) [Ediger et al., 2012] and Ligra [Shun and Blelloch, 2013] (a recent shared-memory parallel library). These offer high performance through native, parallel implementations of certain kernels. For NetworKit to characterize a complex network, it would need a substantial set of analytics, which frameworks currently do not provide. These tools can be further studied for their usability in current regulatory genomic network perspective if efficiency was improved. NetworKit module was preferred in the creation of GraphBreak as it already offers parallel implementations of various community detection algorithms and based on its benchmark performance analysis with closest alternative tools, such as igraph and graph-tool.

As GraphBreak uses NetworKit[Staudt and Meyerhenke, 2015] and NetworkX[Hagberg et al., 2008] Python modules, its benchmarking and comparative benchmarking depends on benchmarking of these modules, which were done by the authors of these tools (cited in the references). NetworKit outperformed igraph and graph-tool in a comparative benchmarking, making it the preference for our 'Big Data' eQTL biological information. Previously, [Itamar Kanter et. al.] used igraph for community detection. Algorithms used for these tools would be similar, but the optimized implementation differs, as that has an effect on the computational resource utilization, parallel compute capability, and other performance metrics, apart from the fact that an easy to use programming interface is also needed.

Another scalability factor is memory footprint of the graph data structure. NetworKit, with 260M edges of the uk-2002 web graph occupies 9 GB with a lean implementation, compared to igraph (93GB) and graph-tool (14GB), benchmarked by the authors NetworKit- increasing our confidence in deploying this module in Python for developing GraphBreak.

Because of NetworKit's competitive disk I/O, the parser is significantly faster for non-attributed graphs. NetworkX was used to load the data to memory for initial calculations and graph plotting then a feature was used to convert toNetworKit format. Although this used additional time for conversion, it was convenient while developing GraphBreak. Comparing NetworKit and the various algorithms implemented by these tools, Figure A, shows that it outperforms others for community detection- making it our preference as a Python module.

NetworKit, igraph and graph-tool rely on hybrid architecture of C/C++ implementations with a Python interface. igraph uses non-parallel C code while graph-tool features parallelism. Graph-tool's approach to community detection is different; hence the comparison is between igraph and NetworKitfor this functionality, which is used in GraphBreak. NetworKit's authors performed the benchmark [Staudt and Meyerhenke, 2015]. GraphBreak would deploy corresponding Python modules to igraph or another community detection tool if drastics improvements were made. NetworKit was the only option for community detection, as closest alternatives, i.e.igraph and graph-tool; require a significantly higher amount of memory and computational time.

I/O should be taken in account for real workflows, as getting a large graph from hard disk to memory often takes longer than actual analysis. For benchmark, authors of NetworKit chose GML graph file format for input files because it is supported by all three frameworks. They observed that NetworKit parser is significantly faster for these non-attributed graphs.

This feature can be considered for future improvements in GraphBreak. Graph data is currently read using NetworkX Python module for initial computation work, such as graph plotting, and uses a converter for NetworKit format for community detection.

GraphBreak was compared to CONDOR [Platig, 2016], results in comparison chart in Fig 10.

| Tool | Principle Underlying Algorithm for Community Detection | Algorithm Implementing Module | Algorithm Implementing Module Relative Performance | Implementation Language | Typical Execution Time for Genomic Associations | Parallel Capability | Automation of Various Steps from data loading to community detection | Community Members Extraction and Ordering functionality | No. of Disease communities detected in Artery Aorta GTEx V7 | No. of Disease communities detected in Artery Coronary GTEx V8 |
|---|---|---|---|---|---|---|---|---|---|---|
| GraphBreak | Louvain Algorithm | networkit | Relatively Faster | Python | Order of minutes | Yes | Yes | Yes | 17 | 8 |
| CONDOR | Michael Barber's Bipartite | igraph | Relatively Slower | R | Order of minutes | No | No | No | 0 | 2 |

**Figure 10**: Relative comparison of GraphBreak and Condor

Broad level comparison of two-regulation based community detection tools, Figure 10, shows GraphBreak can detect a greater number of communities that are over-represented for disease.

**Parallel Mode Choice and Performance**

Speedup would be negatively affected by substantial initial section of the algorithm that is serial in nature. Additionally, speedup would also depend on how well parallelism has been exploited for PLM algorithm. We tested GraphBreak's performance for following system:

Linux 3.10.0-957.12.2.el7.x86_64 #1 SMP Tue May 14 21:24:32 UTC 2019 x86_64 x86_64x86_64 GNU/Linux

for data Artery Coronary significant association in GTEx:

rwxrwxr--. 1 domain users 90M Oct 16 17:10 Artery_Coronary.v8.signif_variant_gene_pairs.txt*

For varying number of compute cores (Parallel Louvain Algorithm) with bipartite plots (Note: this should be done when data is small since bipartite graph plotting takes substantial amount of time, and more importantly, the graph plotted would be uninformative and unaesthetic for large amount of data).

For varying number of compute cores (Parallel Louvain Algorithm) without bipartite plots we see a significant reduction in execution time.

**Results and Discussions**

17 of 56 communities detected using GraphBreak for given dataset of Artery-Aorta eQTL associations for GTEx V7were able to significantly associate its set of genes with disease; for given dataset, 0 of 16 communities detected using Condor showed any statistically significant association. While Condor could predict 2 Artery-Aorta eQTL associations for GTEx V8, GraphBreak was able to predict 8 communities with significant disease association. This illustrates that community detection by different algorithmic approaches lead to different set of results and can have a different impact on downstream analysis. Despite the algorithm being at the disposal of the computer science world, we find GraphBreak novel role in the field of regulation of gene expressions in this gap in research. The data on hand dictates the results of communities. With another dataset, the results from GraphBreak show significant association with diseases and other pathway functionalities while Condor's results underperform.

Many genomic variants were classified based on their regulation information in prior knowledge and could prioritize those which are already known to be an eQTL by some other study for functional validation. They were also investigated for LD. New eQTLs could be found. This paper is software and methods paper, rather than a scientific results paper, so reporting those genes and variants would defeat the purpose. Detailed biological and biomedical significance of results obtained using GraphBreak and Condor for artery-aorta and other relevant tissues for our research interest in coronary artery disease would need to be presented in a separate publication. GraphBreak could aid other scientists with their research interests in other tissues and samples of their interests for GTEx or any other datasets.

**Conclusion and Future Work**

GraphBreak lies at the intersection of network, regulatory genomic science and other analysis such as gene-coexpression. State-of-the-art algorithms for network analysis tasks were incorporated into Python modules, and then coding implementation was done as read-to-use software. This produced a tool suite of network analytics based, community generators and utility software to explore and characterize regulatory networks, such as that of eQTLs network data sets on typical multicore processor.

Scalability with fast parallel algorithms is limited by size of shared memory: a standard multicore workstation with 256 GB RAM can process up to $10^{10}$ edge graphs or functional associations between a genomic variant; this value is large enough even if the p-value for associations between the genomic variant and gene expression is kept high as a threshold. In the future, it could assign different weights per edge connections, such as weight being auto-assigned based on p-value of the associations.

This article shows GraphBreak's application for targeting genes by overexpression analysis and ability to find the corresponding associated genomic regulatory variants. GraphBreak suite could incorporate other community detection algorithms. Although it takes more time than Condor, in its current state GraphBreak performs well for detecting communities. Difference in algorithms used for Condor and GraphBreak impacts community size and members making it unfair to compare the performances of these tools. GraphBreak's actual performance should be benchmarked with its dependency of NetworkX and NetworKit-their authors already showed superior performances in their papers. Newer algorithms could be developed given the NP-hard nature of community detection algorithms, which would be useful in improvising GraphBreak suite to incorporate more algorithmic implementation and options.

From the medical biological perspective, statistical association of community members to any known disease would be valuable. From test data taken for analysis, GraphBreak is better than Condor for this. Many of communities derived from GraphBreak leveraged the importance of community detection for statistically significant disease associations for Artery-Aorta tissue analyzed for GTEx V7, compared to none by Condor; while 2 communities came up with disease association for Artery-Coronary tissue analyzed for GTEx V8 data compared to 8 by GraphBreak. Both can be applied for other relevant tissues for detecting common set of communities having a set of genomic variants regulating a set of gene expression from medical-biological perspective. This paper showed how genes from each community can be analyzed downstream for their statistical associations with disease.

Changing community detection algorithm from Louvain and Girvan-Newman to Leiden algorithm, which may be better than Louvain algorithm as recently published [Traag VA et. al.], could lead to future improvements. As NetworkX module functionality, independent of community detection task, performs graph plotting for bipartite nature of connections, it creates an opportunity of parallelism that can be exploited in future improvisation of GraphBreak code. Future work of GraphBreak would also make a comparison of Girvan-Newman and Leiden algorithm implementation as well for benchmarking purpose. Future development can also see automated downstream analysis of the individual genomic variants such as SNPs in each community with respect to their linkage disequilibrium such as those in cis relation, and also automation to classify the SNPs based on database of pre-existing regulatory information known.

Another possibility of future work would be to use the concept of shingling and hashing as proposed by [Gibson, 2005] to generate graph networks where for each of the nodes the values it contains could be a set of expressed genes, SNPs, known disease, tissue observed, etc., in order to get a consensus of set of genes, SNPs, tissues affected in a disease or trait of interest.

**Open-Source Development and Supplementary**

Diverse community uses and provides contributions through open-source development, including intended community of Bioinformatics professionals. If input data is formatted with right column variable features names, GraphBreak can be used for other network analysis work. The code is free software licensed under the permissive MIT License. Supplementary materials can be downloaded from https://sites.google.com/a/iitdalumni.com/abi/educational-papers .

**Author's contribution**

Abhishek N. Singh developed a research plan to conduct community-based analysis of regulation network. In the process, he developed GraphBreak and executed Condor with exchanges with developers to fix missing features, such as community plot and fix minor bugs. The detailed version of manuscript was done by Abhishek.

**Acknowledgement**

The author acknowledges the makers of CONDOR as inspiration to develop and deploy graph network-based algorithms for biological purposes. The author is thankful to all developers and teachers of Python programming language. Ms Kelin Coleman helped in editing the manuscript.